\documentclass[runningheads]{llncs}
\pdfoutput = 1
\usepackage{enumitem} % Package must be in the preamble
\usepackage[T1]{fontenc}
\usepackage{amsmath}
\usepackage{svg}
\usepackage{caption}
\usepackage{subcaption}
\usepackage{graphicx}
\usepackage{float}
\usepackage{graphicx}
\usepackage{hyperref}
\begin{document}
\title{An Ensemble Approach to Music Source Separation: A Comparative Analysis of Conventional and Hierarchical Stem Separation}
\titlerunning{An Ensemble Approach to Music Source
Separation}
% If the paper title is too long for the running head, you can set
% an abbreviated paper title here
%

\author{
Saarth Vardhan\inst{1}\orcidID{0009-0009-2234-8845}$^{\dagger}$ \and 
Pavani R Acharya\inst{1}\orcidID{0009-0005-2967-9621}$^{\dagger}$ \and 
Samarth S Rao\inst{1}\orcidID{0009-0007-8159-2261}$^{\dagger}$ \and 
Oorjitha Ratna Jasthi\inst{1}\orcidID{0009-0007-0209-5452} \and
S Natarajan\inst{1}\orcidID{0000-0002-8689-5137}
}

\authorrunning{Vardhan et al.}
% First names are abbreviated in the running head.
% If there are more than two authors, 'et al.' is used.

\institute{\textsuperscript{1}Department of Computer Science and Engineering, PES University, Bengaluru, India}
\maketitle              % typeset the header of the contribution
\let\thefootnote\relax\footnotetext{\footnotesize† These authors are equal contributors and share first authorship.}

%
% ---- Abstract ----
%
\begin{abstract}
\noindent Music source separation (MSS) is a task that involves isolating individual sound sources, or stems, from mixed audio signals. This paper presents an ensemble approach to MSS, combining several state-of-the-art architectures to achieve superior separation performance across traditional Vocal, Drum, and Bass (VDB) stems, as well as expanding into second-level hierarchical separation for sub-stems like kick, snare, lead vocals, and background vocals. Our method addresses the limitations of relying on a single model by utilising the complementary strengths of various models, leading to more balanced results across stems. For stem selection, we used the harmonic mean of Signal-to-Noise Ratio (SNR) and Signal-to-Distortion Ratio (SDR), ensuring that extreme values do not skew the results and that both metrics are weighted effectively. In addition to consistently high performance across the VDB stems, we also explored second-level hierarchical separation, revealing important insights into the complexities of MSS and how factors like genre and instrumentation can influence model performance. While the second-level separation results show room for improvement, the ability to isolate sub-stems marks a significant advancement. Our findings pave the way for further research in MSS, particularly in expanding model capabilities beyond VDB and improving niche stem separations such as guitar and piano.
\end{abstract}


\section{Introduction}
Music source separation refers to the process of isolating individual sound sources, or stems, from a mixed audio signal. The primary goal of music source separation is to isolate distinct elements such as vocals, drums, bass, and other instruments from a composite track, enhancing the ability to analyze and work with specific stems.

 Existing architectures have failed to deliver uniformly high performance across the fundamental stems defined by the Vocal-Drums-Bass (VDB) convention. While many models aim to separate these VDB stems effectively, few extend their scope to more complex hierarchies of separation beyond the conventional VDB setup. When such a complex separation is attempted, it typically results in significantly reduced performance across the conventional VDB stems.

As a result, individual architectures, when used in isolation, tend to fall short either in terms of their performance across the VDB stems or in terms of the number or hierarchy of stems they can separate. Furthermore, most of these architectures suffer from a lack of cross-dataset training, which exacerbates the challenges posed by the long tail effect. The long tail effect refers to the phenomenon where certain instruments or stems appear infrequently in training datasets, leading to poorer performance in separating these underrepresented elements.

Commonly used datasets like MUS-DB, DSD100, and MedleyDB\cite{ref_medleydb} contribute to these issues by being individually skewed toward certain genres and instruments, which limits the generalizability of the trained architectures. These datasets also complicate the task of testing second-level hierarchical separation, which is essential for more granular source separation tasks.

Considering all of the above, our research focuses on adeptly combining existing architectures to achieve better overall performance across stems. By leveraging multiple architectures, we can extend the scope of separation beyond the traditional VDB setup to include additional stems such as guitar and piano, as well as enter the second hierarchical level of separation for certain stems.

This combined approach offers several key advantages. First, it consistently achieves uniformly high performance across all conventional VDB stems. Second, it allows us to extend the separation task beyond VDB without compromising the quality of the VDB stems themselves. Additionally, by incorporating multiple architectures, we can adopt a cross-dataset approach, where different datasets, each covering diverse instruments and genres, help to mitigate the long tail effect that affects individual architectures. This leads to improved separation performance for underrepresented instruments, providing a more comprehensive and robust solution to music source separation.

\section{Background}
The earliest work on music source separation (MSS) pertained to separating a track into its four fundamental stems: vocals, drums, bass, and others. As this 4-stem separation approach gained popularity, it was implemented using a variety of architectures, each focusing on different techniques such as mel-spectrograms, raw audio waveforms, unsupervised learning from unlabelled data (particularly in scenarios of low data availability), recurrent neural networks\cite{ref_dual_path_rnn}, specific frequency bands, generative adversarial networks (GANs)\cite{ref_gan_unet}, transformers, convolutional neural networks (CNNs)\cite{ref_audio_stems}, multitask learning, and DTT-Net\cite{ref_dttnet}.

A rudimentary analysis of certain architectures, including hybrid transformers, rotary positional encoding transformers\cite{ref_mel_band_roformer}, sparse compression networks\cite{ref_scnet}, and adversarial networks\cite{ref_adversarial_separation}, revealed higher performance on specific stems. This superior performance could largely be attributed to one common feature across these architectures: attention mechanisms.

Attention mechanisms allow these architectures to dynamically focus on specific time-frequency regions within audio data, enhancing the separation of overlapping sources. By selectively attending to relevant features such as vocal harmonics or rhythmic patterns, attention-driven architectures achieve higher accuracy in isolating complex musical components like vocals, drums, and bass.

Following these developments, datasets were curated to include stems beyond the traditional VDBO (vocals, drums, bass, others) structure, with guitar and piano serving as key additions to this research. Given the prominence of guitar and piano in western music, MSS research expanded to incorporate these stems into the separation models. This was then further scaled to the second hierarchical level, where stems within primary stems, such as drums, vocals and guitar, were considered for separation.

\section{Related Work}
Recent advancements in music source separation (MSS) have largely been driven by deep learning models, which have demonstrated significant improvements over traditional approaches. These models employ a range of innovative techniques, each with its own focus on enhancing source separation performance. Below, we explore several key models that have contributed to the current state of the field.

\subsection{SCNet (Sparse Compression Network)}
SCNet, proposed by Tong et al.\cite{ref_scnet}, is a frequency-domain network that splits spectrograms into subbands. Its key innovation is a sparse compression technique that applies higher compression ratios to less informative subbands, maintaining fidelity in critical subbands while reducing computational load. This approach enhances separation performance and reduces model complexity, making SCNet suitable for real-time applications. Its efficient handling of subbands and sparse compression make SCNet stand out in terms of both performance and efficiency.

\subsection{Mel-Band RoFormer}
The Mel-Band RoFormer, introduced by Wang et al.\cite{ref_mel_band_roformer}, extends BS-RoFormer\cite{ref_band_split_rope} by using a mel-scale-based projection scheme to capture perceptual differences in frequencies, particularly in vocals and drums. Its key innovation is the use of overlapping subbands based on the mel scale, mimicking the human auditory system’s preference for higher resolution at lower frequencies. This improves the separation of stems sensitive to frequency resolution, such as vocals and drums. Its hierarchical Transformer with Rotary Position Embedding (RoPE) enhances both inner-band and inter-band dependency modeling, making it highly effective for multi-band mask estimation.

\subsection{Hybrid Transformer Demucs (HT Demucs)}
The Hybrid Transformer Demucs (HT Demucs), developed by Rouard et al., \cite{ref_demucs}combines temporal and spectral U-Nets with Transformer layers to enhance its receptive field. By integrating both time and frequency domains, HT Demucs applies self-attention within and cross-attention across them, capturing short- and long-term dependencies for complex musical mixtures. Its Transformer-based approach, with sparse attention kernels, extends the receptive field while keeping computational costs low. Fine-tuned on an extensive dataset. This model’s hybrid structure makes it highly effective for MSS.

\subsection{Band-Split RNN for MSS}
Luo and Yu propose Band-Split RNN (BSRNN)\cite{ref_band_split_rnn}, a frequency-domain MSS model that explicitly divides the input spectrogram into subbands based on predefined bandwidths. The model operates in the high-sample-rate frequency domain, allowing it to handle super wide-band signals efficiently. It leverages both sequence-level and band-level processing via RNNs to capture intra-band dependencies. BSRNN is designed to improve performance by focusing on the most relevant frequency components for each instrument, such as lower frequencies for bass.

\subsection{Wave-U-Net: End-to-End Separation in the Time Domain}
\noindent Wave-U-Net, proposed by Stoller et al.\cite{ref_wave_u_net}, is a time-domain adaptation of the U-Net architecture designed for audio source separation. Unlike spectrogram-based methods, Wave-U-Net operates directly on waveforms, enabling the model to process both magnitude and phase information without relying on fixed transformations. It uses multi-scale feature maps to capture long-term temporal dependencies, essential for accurate separation. Additionally, it introduces a context-aware prediction framework and a refined upsampling technique to minimize artifacts. The model demonstrated competitive performance, particularly in singing voice separation, against state-of-the-art spectrogram-based systems.

\section{Proposed Methodology}
\subsection{Test Set Description}
In this research, we curated a test set comprising 71 songs from the Moises-DB dataset. Moises-DB\cite{ref_moisesdb}, widely regarded for its comprehensive and well-annotated multitrack datasets, is particularly suited for our study due to its wide range of musical genres and the presence of meticulously labeled stems extending to the second hierarchy. These characteristics make it ideal for evaluating separation algorithms, as it provides a representative benchmark for our models.

\begin{figure}[htbp]
    \centering
    \includegraphics[width=0.95\linewidth]{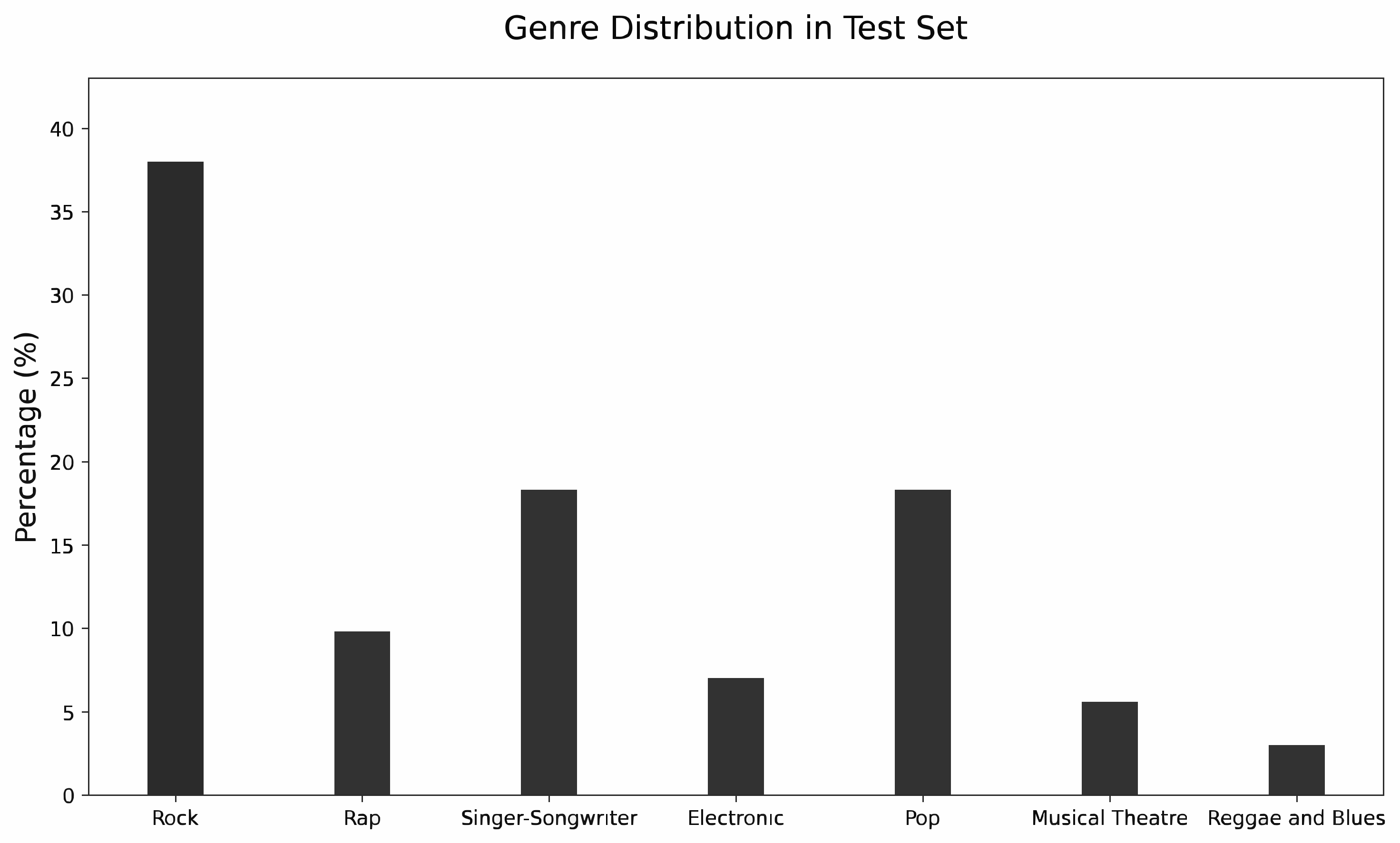}\hfill
    \caption{Genre Distribution in Test Set}
    \label{fig:genre_distribution}
\end{figure}

This test set provided ground truth stems for both conventional VDB and second hierarchical level stems. The test set spans a broad range of genres, ensuring diverse audio challenges and is representative of the entire Moises-DB dataset. Rock is the most represented genre at 38\%, followed by Singer-Songwriter and Pop, each at 18.3\%. Rap accounts for 9.8\%, while Electronic adds 7\%. Lastly, Musical Theatre and Reggae and Blues contribute 5.6\% and 3\%, respectively. This distribution provides a comprehensive mix of styles, ideal for evaluating our separation models, as shown in \textbf{Figure~\ref{fig:genre_distribution}}.
\vspace{-0.75cm}
\subsection{For VDB Stems}

As highlighted in the previous sections, the primary objective for the VDB (Vocals, Drums, Bass) stems is to achieve consistently high separation performance across all stems. This has been one of the most enduring challenges in MSS, particularly due to the diverse nature of music signals and their complex interdependencies. To tackle this, we utilized an ensemble of state-of-the-art architectures, each specializing in different aspects of the separation process. This ensemble primarily consisted of hybrid transformers, RoPE (Rotary Position Embedding) transformers, and sparse compression networks. These models, although effective in isolation, complement each other when used together, addressing specific limitations and enhancing overall performance.

The ensemble approach utilises the strengths of each architecture. Hybrid transformers\cite{ref_demucs,ref_github2}, known for their ability to model both long-range and short-range dependencies, are well-suited for capturing temporal and frequency variations in musical signals. RoPE transformers\cite{ref_mel_band_roformer,ref_github1}, with their positional encoding in complex planes, allow for more fine-grained separation, particularly in challenging audio scenarios where frequency overlap is common. Sparse compression networks\cite{ref_scnet,ref_github3}, on the other hand, enhance computational efficiency by focusing on regions of the spectrogram that contain the most critical information, reducing the load on less informative subbands. By integrating these diverse architectures, we ensured that the ensemble worked synergistically to handle various song genres and musical structures, resulting in a higher overall separation quality than could be achieved by any single architecture.

In order to utilise these architectures for our research, we made use of open-source model weights and configuration files. For each track in our test set, the separation process was conducted individually using each architecture. After separation, both Signal-to-Noise Ratio (SNR) \ref{eq:snr} and Signal-to-Distortion Ratio (SDR) \ref{eq:sdr} were computed for each stem across all models. These metrics were computed against the ground truth stem files obtained from Moises-DB\cite{ref_moisesdb}, ensuring a rigorous quantitative comparison across all stems and models.

The next step involved aggregating the outputs from the various models to determine the best-quality stem for each song and each stem type. A bootstrapped aggregation technique was employed for this purpose, which not only ensures robustness in the selection process but also reduces the risk of overfitting to specific models. This technique was particularly useful in avoiding the pitfalls of selecting stems based on a single model’s performance, which might excel in one aspect but falter in another.

To aggregate results effectively, we needed a reliable comparison metric. In audio source separation, SNR and SDR are the most commonly used metrics due to their ability to reflect both signal fidelity and noise suppression. However, simple arithmetic mean calculations of SNR and SDR are often insufficient for selecting the best-quality stem. The arithmetic mean tends to give equal weight to both SNR and SDR, which can be problematic when these values differ significantly, leading to skewed results. Moreover, arithmetic means are more sensitive to outliers, which can disproportionately affect the final stem quality.

Our research identified the harmonic mean as a more appropriate metric for this task. The harmonic mean emphasizes the smaller of the two values (SNR and SDR), ensuring that stems with poor performance in either metric do not dominate the final selection. This is critical because high SNR alone may not be indicative of good separation quality if the SDR is low, and vice versa. By using the harmonic mean, we penalize outlier values more effectively and ensure that the selected stems perform well in both metrics, striking a balance between signal clarity and distortion reduction. Additionally, the harmonic mean is less sensitive to extreme values, which provides a more stable and reliable method for stem selection in the presence of varying signal qualities across different models.

Thus, the model output with the highest harmonic mean for each stem was chosen as the "best" stem for that track. The final results were computed, and the model used for each stem in each song was recorded for analysis. This process enabled us to achieve consistently high performance across all VDB stems (vocals, drums, and bass) throughout the test set, reducing reliance on any single model and promoting a more generalized approach to MSS.
\begin{equation}
    \text{SNR} = 10 \log_{10}\left(\frac{P_{\text{signal}}}{P_{\text{noise}}}\right)
    \label{eq:snr}
\end{equation}

where \( P_{\text{signal}} \) is the power of the signal and \( P_{\text{noise}} \) is the power of the noise.\\

\begin{equation}
    \text{SDR} = 10 \log_{10}\left(\frac{||x||^2}{||x - \hat{x}||^2}\right)
    \label{eq:sdr}
\end{equation}

where \( x \) is the original signal and \( \hat{x} \) is the estimated signal.
\subsection{Second-Level Hierarchical Separation}
The long-term applicability of MSS is greatly enhanced by expanding its capabilities to handle more fine-grained separations within primary stems, a process we refer to as "second hierarchical separation". This involves separating broad categories, such as vocals or drums, into their constituent parts—vocals into lead male/female and background vocals, and drums into toms, cymbals, kick drum, and snare drum.

Surprisingly, very few architectures have ventured into second hierarchical separation, primarily due to the lack of a widely available, well-annotated dataset capable of supporting this type of task. Additionally, the acoustic similarities between sub-stems, the frequency overlap, and the complex mix interactions (such as harmonics and spatial blending) present significant challenges. Most existing models are designed to handle broader stem separation and struggle with the intricate distinctions required for second-level separation.

Despite these challenges, second hierarchical separation is critical for applications where detailed stem control is needed, and its importance cannot be overstated. While this separation may come at the cost of slightly decreased performance in some areas (due to the inherent complexity), the ability to isolate these sub-stems alongside the VDB stems is of immense value both from a research perspective and for practical applications in music editing, restoration, and content creation.

Our research focuses on utilizing a RoPE-Transformer based architecture\cite{ref_band_split_rope,ref_github5} for the separation of vocal stems into lead male/female vocals and background vocals. RoPE's ability to model fine-grained dependencies within time-frequency representations makes it an ideal candidate for this task, particularly in distinguishing subtle vocal harmonies and overlapping frequencies. Additionally, we employ a Demucs-based architecture for drum stem separation. Demucs\cite{ref_demucs,ref_github4}, known for its strong performance in time-domain signal separation, is particularly well-suited for handling the complex interactions within drum stems, separating toms, kick drums, snare drums, and cymbals.

In conclusion, our approach not only achieves a consistently high performance across conventional VDB stems but also ventures into the second hierarchical level of separation for crucial stems like drums and vocals. This dual focus addresses both broad and fine-grained separation needs, making our research a significant step forward in the field of music source separation.

\begin{figure}[htbp]
    \centering
    \includegraphics[width=1\linewidth,height=2cm]{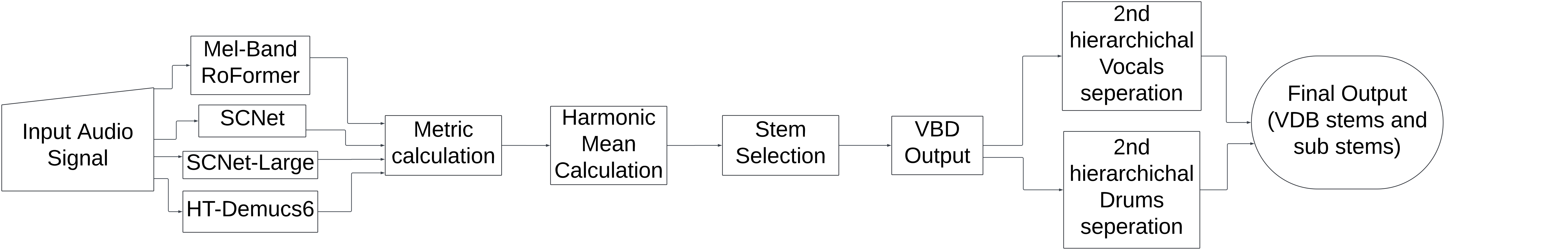}
    \caption{Diagrammatic representation of our approach}
    \label{fig:flow_diagram}
\end{figure}
\vspace{-1cm}
\section{Results}
Our ensemble was tested with \textbf{four} primary models mentioned below.\\ \\
\noindent \textit{MelBand RoFormer}\cite{ref_mel_band_roformer}. Performs vocals-only separation\cite{ref_github1}.\\ \\
\textit{SC-Net}\cite{ref_scnet}. Performs separation of vocals, bass, and drums (VDB)\cite{ref_github3}.\\ \\
\textit{SC-Net Large}\cite{ref_scnet}. A more advanced variant of SC-Net trained on a larger dataset, also used for VDB separation\cite{ref_github3}.\\ \\
\textit{HT-Demucs6}\cite{ref_demucs}. Capable of separating not only VDB stems but also the piano and guitar stems\cite{ref_github2}.
\subsection{Performance of Individual Models on VDB Stems}
\begin{figure}[htbp]
    \centering
    \begin{subfigure}[t]{0.48\textwidth}
        \centering
        \includegraphics[width=\linewidth]{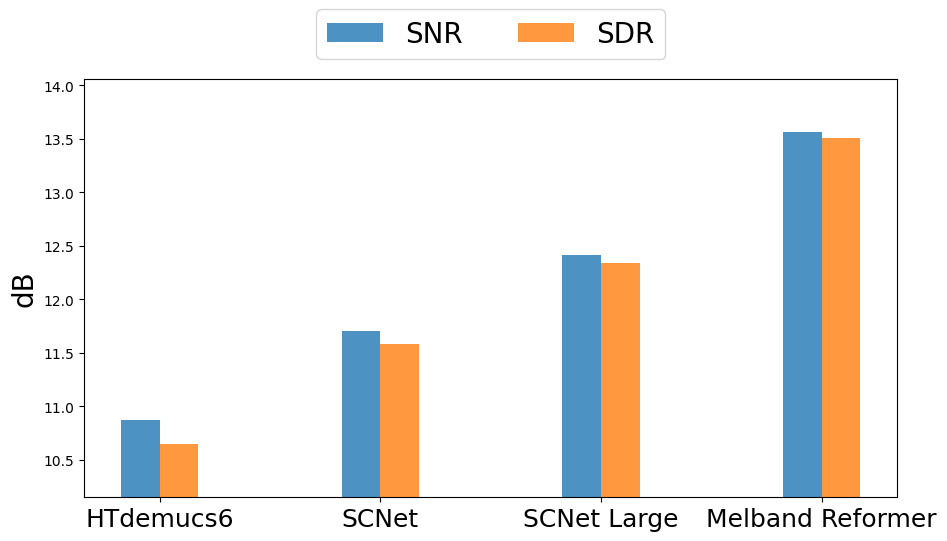}
        \caption{\scriptsize SNR and SDR for Vocals}
        \label{fig:snr_sdr_vocals}
    \end{subfigure}%
    \hspace{0.03\textwidth} % Adjust spacing between figures
    \begin{subfigure}[t]{0.48\textwidth}
        \centering
        \includegraphics[width=\linewidth]{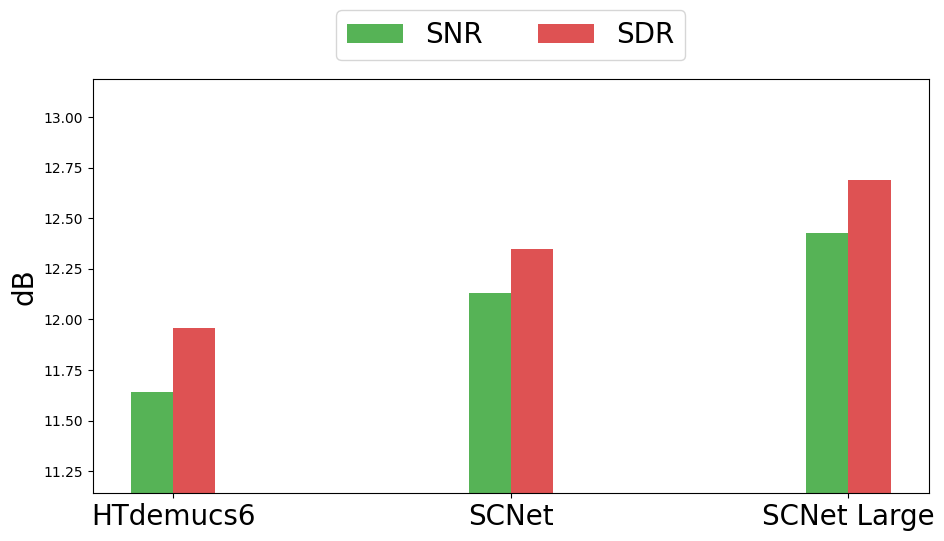}
        \caption{\scriptsize SNR and SDR for Bass}
        \label{fig:snr_sdr_bass}
    \end{subfigure}
    
    \vspace{0.03\textwidth} % Adjust vertical spacing between rows
    
    \begin{subfigure}[t]{0.48\textwidth}
        \centering
            \includegraphics[width=\linewidth]{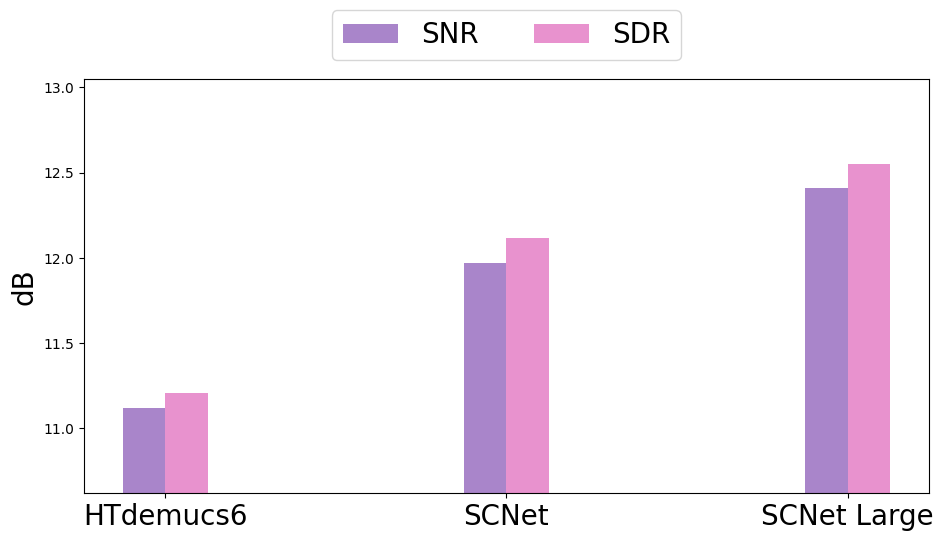}
        \caption{\scriptsize SNR and SDR for Drums}
        \label{fig:snr_sdr_drums}
    \end{subfigure}%
    \hspace{0.03\textwidth} % Adjust spacing between figures
    \begin{subfigure}[t]{0.48\textwidth}
        \centering
        \includegraphics[width=\linewidth]{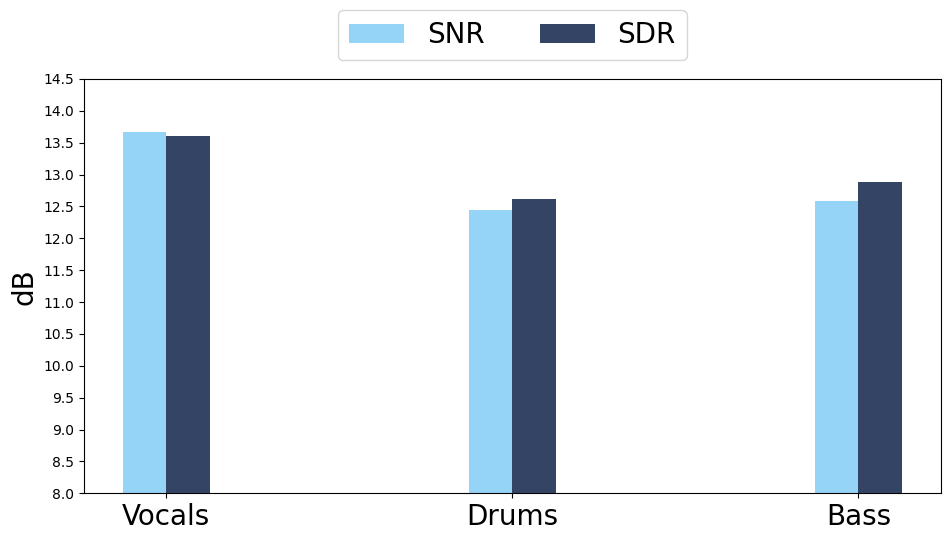}
        \caption{\scriptsize SNR and SDR for VDB of Final Ensemble}
        \label{fig:vdb_snr_sdr}
    \end{subfigure}
    
    \caption{SNR and SDR Comparison across Different Models and our Final Ensemble for Vocals, Bass and Drums}
    \label{fig:snr_sdr_square}
\end{figure}

After running all models on the 71-song test set, we computed the SNR and SDR for each stem (vocals, drums, bass, guitar, piano). \textbf{Table}\textbf{~\ref{tab:snr_sdr_vdb}} summarizes the average results across the different models.
\noindent \textbf{Figures~\ref{fig:snr_sdr_vocals}}, \textbf{~\ref{fig:snr_sdr_bass} and \ref{fig:snr_sdr_drums}} shows the breakdown of SNR and SDR for vocal, bass and drums separation respectively across the four models. The results of our ensemble can be inferred from \textbf{Table}\textbf{~\ref{tab:snr_sdr_vdb}} and \textbf{Figure~\ref{fig:vdb_snr_sdr}}
\vspace{-0.8cm}
\begin{table}[H] % Use [H] to force placement here
\centering
\caption{SNR and SDR Comparison across Models}
\label{tab:snr_sdr_vdb}
\vspace{0.1cm} % Adjust the value to control the space
\setlength{\tabcolsep}{1pt} % Increase column spacing
\renewcommand{\arraystretch}{1.3} % Increase row spacing
\begin{tabular}{|c|c|c|c|c|c|c|c|c|c|c|}
\hline
\textbf{Model}          & \multicolumn{2}{c|}{\textbf{Vocals}} & \multicolumn{2}{c|}{\textbf{Drums}} & \multicolumn{2}{c|}{\textbf{Bass}} & \multicolumn{2}{c|}{\textbf{Guitar}} & \multicolumn{2}{c|}{\textbf{Piano}} \\ 
\cline{2-11}
                        & \scriptsize\textbf{SNR} & \scriptsize\textbf{SDR} & \scriptsize\textbf{SNR} & \scriptsize\textbf{SDR} & \scriptsize\textbf{SNR} & \scriptsize\textbf{SDR} & \scriptsize\textbf{SNR} & \scriptsize\textbf{SDR} & \scriptsize\textbf{SNR} & \scriptsize\textbf{SDR} \\ \hline
HT-Demucs6\cite{ref_github2}              & 10.87        & 10.65        & 11.12        & 11.21        & 11.64        & 11.96        & 4.38         & 2.59         & 1.62         & -6.25        \\ \hline
MelBand RoFormer\cite{ref_github1}        & 13.56        & 13.51        & -            & -            & -            & -            & -            & -            & -            & -            \\ \hline
SC-Net\cite{ref_github3}                  & 11.71        & 11.58        & 11.97        & 12.12        & 12.13        & 12.35        & -            & -            & -            & -            \\ \hline
SC-Net Large\cite{ref_github3}            & 12.41        & 12.34        & 12.41        & 12.55        & 12.43        & 12.69        & -            & -            & -            & -            \\ \hline
\textbf{Final Ensemble}           & \textbf{13.66}        & \textbf{13.61}        & \textbf{12.45}        & \textbf{12.61}        & \textbf{12.59}       & \textbf{12.88}        & \textbf{4.38}            & \textbf{2.59}            & \textbf{1.62}           & \textbf{-6.25}            \\ \hline
\end{tabular}
\end{table}
\vspace{-0.5cm}
\subsection{Second Hierarchical Separation}
For second hierarchical separation, \textbf{Drumsep}\cite{ref_github4} was used for drum separation into sub-stems, and a variant of \textbf{MelBand RoFormer}\cite{ref_github5} was used for vocal sub-stems. \textbf{Tables~\ref{tab:snr_sdr_drums_2nd} and \ref{tab:snr_sdr_vocals_2nd}} present the SNR and SDR for the sub-stems.
\vspace{-0.5cm}
\captionsetup{font=small}
\begin{table}[htbp]
\centering
\begin{minipage}[t]{0.48\textwidth}
    \centering
    \caption{2nd hierarchical drum separation}
    \label{tab:snr_sdr_drums_2nd}
    \vspace{0.1cm} % Adjust the value to control the space
    \setlength{\tabcolsep}{6pt} % Increase column spacing
    \begin{tabular}{|c|c|c|}
    \hline
    \textbf{Drum Sub-Stem} & \textbf{SNR} & \textbf{SDR} \\ \hline
    Kick Drum              & 12.87        & 13.65        \\ \hline
    Snare Drum             & 7.26         & 7.52         \\ \hline
    Toms                   & 4.60         & 3.26         \\ \hline
    Cymbals                & -2.98        & -5.64        \\ \hline
    \end{tabular}
\end{minipage}%
\hfill
\begin{minipage}[t]{0.48\textwidth}
    \centering
    \caption{2nd hierarchical vocal separation}
    \captionsetup{font=tiny} % Change the font size here
    \label{tab:snr_sdr_vocals_2nd}
    \vspace{0.1cm} % Adjust the value to control the space
    \setlength{\tabcolsep}{6pt} % Increase column spacing
    \renewcommand{\arraystretch}{1.25} % Increase row spacing
    \begin{tabular}{|c|c|c|}
    \hline
    \textbf{Vocal Sub-Stem} & \textbf{SNR} & \textbf{SDR} \\ \hline
    Lead Vocal (Male)       & -0.73        & 10.34        \\ \hline
    Lead Vocal (Female)     & -1.39        & 11.86        \\ \hline
    Background Vocal        & -7.57        & -0.80        \\ \hline
    \end{tabular}
\end{minipage}
\end{table}
\vspace{-0.5cm}
\subsection{Discussion}
\noindent As previously outlined in Table~\ref{tab:snr_sdr_vdb}, our final ensemble solution consistently exhibits superior performance across the VDB stems (vocals, drums, bass) compared to individual models. By combining multiple architectures and utilising their respective strengths, we observe substantial improvements in both SNR and SDR across all key stems. As we can see in Table~\ref{tab:snr_sdr_vdb}, our final solution shows a superior performance in vocal separation, achieving an average SNR of 13.66 dB, which outperforms the best performing model for vocal separation - MelBand RoFormer. Similarly, for bass and drum separation as well, the proposed solution achieves an SNR of 12.59 and 12.45 dB respectively, which is a clear improvement in performance from the existing models that perform bass and drum separation. This highlights the uniqueness of our approach, which integrates multiple models rather than relying on a single architecture, allowing us to outperform any individual existing architecture, as discussed in the Proposed Methodology section.

The results of our foray into second hierarchical separation, as shown in Tables~\ref{tab:snr_sdr_drums_2nd} and \ref{tab:snr_sdr_vocals_2nd}, further demonstrate the robustness of our approach. While separating complex sub-stems (such as lead and background vocals, or toms and cymbals in drums) presents inherent challenges, the outcomes are promising. The harmonic mean of SNR and SDR across the dataset shows that our ensemble maintains strong performance despite these difficulties. This ties back to the discussion on model aggregation in the Proposed Methodolody section, where the harmonic mean was selected specifically to handle disparate performances across different models, ensuring consistently high-quality outputs.

\begin{figure}[H]
    \centering
    \includegraphics[width=0.65\linewidth]{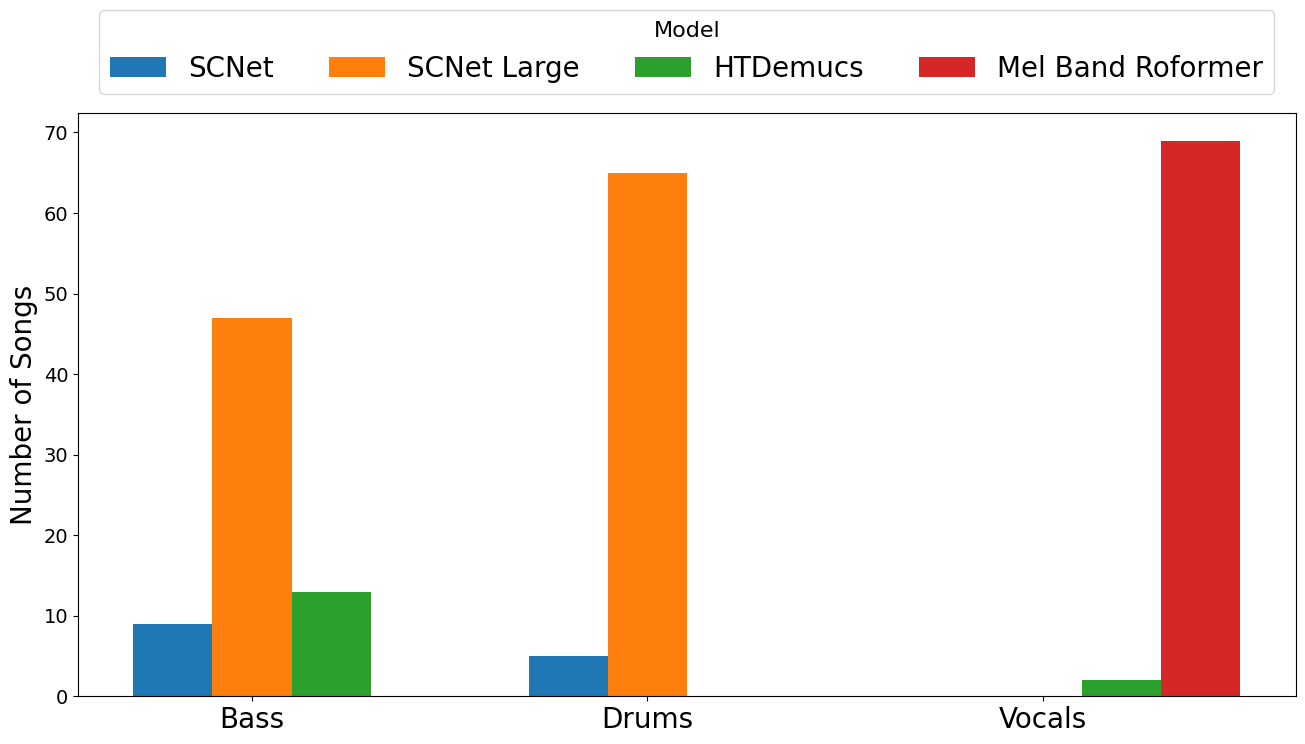}\hfill
    \caption{Number of Songs by Model and Stem}
    \label{fig:songs_by_model_stem}
\end{figure}

As seen in \textbf{Figure~\ref{fig:songs_by_model_stem}}, key observations from our experiments are mentioned below.\\ \\
\textbf{MelBand RoFormer}\cite{ref_mel_band_roformer}\textbf{.} Dominates vocal separation, being selected for 97\% of the songs, consistently producing higher-quality vocal stems.\\

\noindent \textbf{SCNet Large}\cite{ref_scnet}\textbf{.} Stands out as the top performer for bass separation, selected in 67\% of the cases, indicating its superior ability to capture low-frequency nuances. HTDemucs\cite{ref_demucs} and SCNet\cite{ref_scnet} were chosen less frequently, showing that they are less reliable for bass separation.\\ 

\noindent \textbf{SCNet Large}\cite{ref_scnet}\textbf{.} This model was also preferred for drum separation in 93\% of the cases, confirming its effectiveness in handling complex transients and delivering clean drum stems.\\
\vspace{-0.75cm}
\section{Analysis}
\vspace{0.1cm}
\subsection{Correlation Between Model Performance, Stems, and Genres}
We observed certain correlations between model performance and specific genres which are mentioned below.\\ \\
\textbf{Vocals.} In genres like pop, musical theatre, and blues, where vocals are prominent and well-mixed, models consistently achieved higher SNR/SDR values. The clarity and emphasis on vocals in these genres made them easier for models to isolate.\\
    
    \noindent\textbf{Bass.} In electronic, rap, and reggae, where bass plays a central role, models had a harder time isolating bass stems due to the overlap with kick drums and other low-frequency elements. This is reflected in the lower SNR/SDR values for bass in these genres.\\
    
    \noindent\textbf{Percussion.} Drum stems were more challenging to separate in genres like rock and electronic, where complex effects (distortion, reverb) and heavy production techniques created difficulties in isolating individual percussion components.
\subsection{Model-Specific Analysis}
Each model in our ensemble was designed with unique objectives in mind, and these differences shape their performance across different instrument stems. The contribution of each architecture to the overall result is mentioned below. \\

    \noindent\textbf{SCNet (Sparse Compression Network)}\cite{ref_scnet,ref_github3}\textbf{.} SCNet’s frequency-domain approach, which splits spectrograms into subbands and applies a sparsity-based encoder, is particularly well-suited for separating sustained low-frequency sounds such as bass. Its ability to focus on spectrogram regions that carry the most relevant information explains its success in bass separation. The SCNet Large variant, with its increased model capacity (more layers and parameters), was able to handle complex tasks like bass and drum separation more effectively, providing the best results in terms of SNR and SDR.\\
    
    \noindent \textbf{Mel-Band RoFormer}\cite{ref_mel_band_roformer,ref_github1}\textbf{.}This model excels in vocal separation because of its mel-scale projection, which mimics human auditory perception by focusing on mid- and high-frequency ranges where vocals typically reside. The overlapping subbands help to maintain vocal clarity while filtering out interference from other instruments. Its dominance in the vocal stem (97\%) underscores its strength in this domain and highlights its specific design advantages.\\

    \noindent\textbf{HT Demucs}\cite{ref_demucs,ref_github2}\textbf{.} HT Demucs, though not the leading model for any specific VDB stem, excels in versatility. Its hybrid approach (combining time and frequency domains with attention mechanisms) makes it adaptable to multiple types of stems. HT Demucs performs well in scenarios with complex mixtures, particularly in the piano and guitar stems, where a hybrid representation is beneficial.

\subsection{Factors Influencing Model Performance}
\begin{figure}[htbp]
    \centering
    \includegraphics[width=0.75\linewidth]{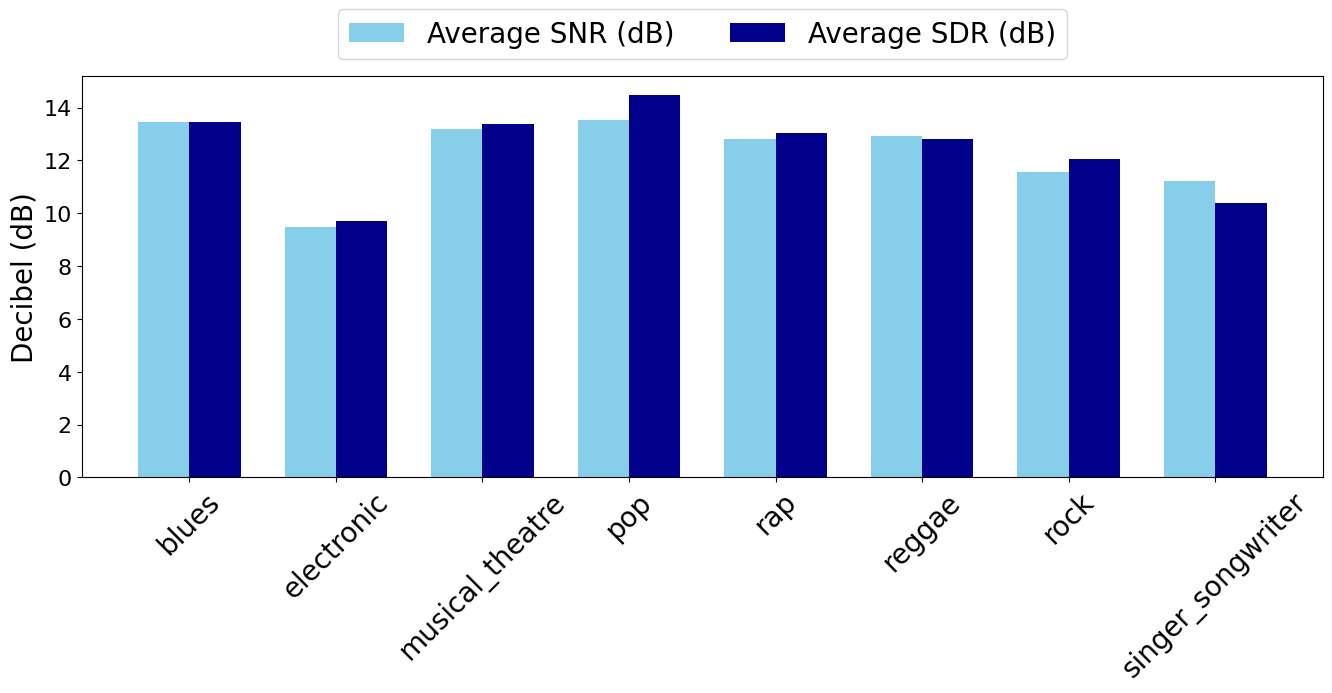}
    \caption{Average SNR and SDR by Genre}
    \label{fig:snr_sdr_by_genre}
\end{figure}

As seen in \textbf{Figure~\ref{fig:snr_sdr_by_genre}}, the genre of the track plays a significant role in determining how well a model performs on specific stems. Our inferences on how genre-related factors influenced SNR and SDR in our study are detailed below.\\

    \noindent\textbf{Instrumentation Complexity}\textbf{.} Simpler genres, such as blues and pop, which feature structured instrumentation, tend to achieve higher SNR and SDR. Complex genres like electronic and rock, with overlapping frequencies and heavy processing, pose greater challenges for separation models, leading to lower metric values.\\
    
    \noindent\textbf{Frequency and Harmonic Content}\textbf{.} In genres like electronic and rap, which rely on low-frequency elements (bass, sub-bass), there is significant overlap between stems like bass and drums, making separation challenging. On the other hand, genres like musical theatre, singer-songwriter, blues, and pop, which emphasize clean and prominent vocals, show better separation performance due to easier isolation of vocal harmonics. As a result, these genres typically achieved higher SNR and SDR values for vocal separation.
\vspace{-0.5cm}
\subsection{Second Hierarchical Separation of Vocals and Drums}
The foray into second hierarchical separation, particularly for vocals and drums, was a challenging but crucial part of the research. The insights from our analysis of SNR and SDR for each vocal and drum component are highlighted below.\\

\noindent\textbf{Vocals.} Interestingly, the has\_bleed parameter in our dataset provided insight into the effect of audio bleed on vocal separation. For lead vocals, the difference in Mod(SNR-SDR) between songs with and without bleed was negligible (0.04 dB), suggesting that bleed has minimal impact on lead vocal separation. This may be due to the centrality of lead vocals in most mixes and the care taken during recording and post-processing to mitigate bleed.\\
    
    \noindent \textbf{Background vocals.} However, the Mod(SNR-SDR) difference was more significant (2.82 dB) for background vocals, with songs featuring bleed showing lower scores. This implies that background vocals are more susceptible to bleed, potentially blending with other stems, which complicates clean separation.\\
    
    \noindent \textbf{Drums.} The performance for drum separation components was varied. Kick drum separation achieved the highest SNR and SDR values, likely due to its distinct low-frequency presence and minimal overlap with other components. In contrast, snare drum performance was moderate, as bleed and mid-frequency overlap reduced its separation quality. Toms and cymbals had the lowest separation quality, with toms presenting challenges due to their broader frequency range and cymbals proving difficult because of their complex high-frequency content and bleed across multiple microphones, making separation more challenging.
    
\section{Conclusion and Future Work}

\noindent In this study, we demonstrated that an ensemble of existing architectures can consistently outperform any individual model in the separation of traditional VDB stems. Additionally, we explored second-level separation for drums and vocals. While the ability to separate these finer components is a significant achievement using current architectures, there is still room for improvement in the overall performance of these second-level tasks. Nonetheless, this exploration has provided valuable insights into the complexities of the source separation process and highlighted how factors like genre and instrumentation can influence model performance. 

While our research has advanced the field of music source separation, several areas remain open for further development. One potential avenue is to expand current VDB-focused models to also encompass instruments such as guitar and piano, broadening their scope and enhancing their application to more diverse musical compositions. Additionally, while we have made progress in second hierarchical separation, future work could aim to improve the overall performance of models handling this task, with the goal of achieving greater accuracy and separation quality for more granular instrument components. Lastly, there is an opportunity to develop individual models that focus on more niche stems, such as violin or other orchestral instruments, which are often underrepresented in existing models. By exploring these areas, future research can build upon the foundation we have established, creating more versatile and precise music source separation models. Our findings thus emphasize the potential of ensemble approaches in music source separation and pave the way for future work that can further refine these methods.

\subsubsection{\ackname} 
We would like to thank the developers who generously shared their models and resources as open source, making this work possible. Special thanks go to Kimberley Jensen for the Mel-Band Roformer Vocal Model, the Facebook AI Research team for Demucs, and the creators of SCNet for their essential contributions. Inagoy’s Drumsep also played a key role in this research. Additionally, appreciation goes to the contributors behind the Python Audio Separator, including Anjok07, DilanBoskan, Kuielab, Woosung Choi, Hv, and zhzhongshi, for their work on vocal separation techniques for 2nd level vocal seperation, which significantly enhanced our research's capabilities.

\subsubsection{\discintname}
The authors have no competing interests to declare that are relevant to the content of this article.

\newpage
% Include the natbib package for square bracket citations

% ---- Bibliography ----
% BibTeX users should specify bibliography style 'splncs04'.
% References will then be sorted and formatted in the correct style.
%
% \bibliographystyle{splncs04}
% \bibliography{mybibliography}

% ---- Bibliography ----

\end{document}